\documentclass[article]{aa}
\usepackage{graphicx}
\usepackage[varg]{txfonts}
\begin{document}

\title{Deriving physical parameters of unresolved star clusters}
\subtitle{II. The degeneracies of age, mass, extinction, and metallicity}
\author{P. de Meulenaer\inst{1,2} \and D. Narbutis\inst{2} \and T. Mineikis\inst{1,2} \and V. Vansevi\v{c}ius\inst{1,2}}

\institute{Vilnius University Observatory, \v{C}iurlionio 29, Vilnius LT-03100, Lithuania\\
\email{philippe.demeulenaer@ff.stud.vu.lt} \and Center for Physical Sciences and Technology, Savanori\c{u} 231, Vilnius LT-02300, Lithuania}

\date{Submitted April 14, 2014; Accepted June 26, 2014}

\abstract
{This paper is the second of a series that investigates the stochasticity and degeneracy problems that hinder the derivation of the age, mass, extinction, and metallicity of unresolved star clusters in external galaxies when broad-band photometry is used.}
{While Paper I concentrated on deriving age, mass, and extinction of star clusters for one fixed metallicity, we here derive these parameters in case when metallicity is let free to vary. The results were obtained using several different filter systems ($UBVRI$, $UBVRIJHK$, GALEX+$UBVRI$), which allowed to optimally reduce the different degeneracies between the cluster physical parameters.}
{The age, mass, and extinction of a sample of artificial star clusters were derived by comparing their broad-band integrated magnitudes with the magnitudes of a large grid of cluster models with various metallicities.}
{A large collection of artificial clusters was studied to model the different degeneracies in the age, mass, extinction, and metallicity parameter space when stochasticity is taken into account in the cluster models. We show that, without prior knowledge on the metallicity, the optical bands ($UBVRI$) fail to allow a correct derivation of the age, mass, and extinction because of the strong degeneracies between models of different metallicities. Adding near-infrared information ($UBVRI$+$JHK$) slightly helps in improving the parameter derivation, except for the metallicity. Adding ultraviolet data (GALEX+$UBVRI$) helps significantly in deriving these parameters and allows constraining the metallicity when the photometric errors have a Gaussian distribution with standard deviations 0.05 mag for $UBVRI$ and 0.15 mag for the GALEX bands.}
{}


\maketitle

\section{Introduction}
Star clusters are essential tools for understanding the formation and evolution of their host galaxies. Indeed, it is commonly assumed that a large fraction of stars were born in star clusters or stellar associations \citep{Lada2003}. Know the physical parameters of the star cluster population (i.e., age, mass, chemical composition, extinction) is essential for determining of star formation history in a galaxy, and can be used as constraints for galaxy evolution models.

The traditional photometric method used to derive the star cluster physical parameters is based on the comparison of the predicted integrated colors of a set of simple stellar population (SSP) models with the integrated broad-band photometry of the observed clusters, as reported \citet{Anders2004} and \citet{Bridzius2008}. However, this method is strongly biased by degeneracies between parameters and by the stochastic presence of bright stars that dominate the integrated photometry. The problem of deriving of the age and mass of clusters taking stochasticity into account has been studied in, for instance, by \citet{Popescu2009, Popescu2010}, when the photometric data were corrected for extinction before the analysis. Simultaneous derivation of age, mass, and extinction has been studied in \citet{Fouesneau2010}, \citet{Fouesneau2012}, \citet{de_Meulenaer2013} (hereafter Paper I), and in \citet{Anders2013}. A recent review of the progress in the field has been compiled by \citet{Cervino2013}. \citet{Anders2013} also studied the derivation of metallicity for narrow ranges of extinction and metallicity search, centered on the input values. The aim of this paper is to extend the study of \citet{de_Meulenaer2013} to a wide range of metallicities, from $Z=0.03$ to $Z=0.00013$. We also present the ability of the method to derive the star cluster parameters using different filter sets: $UBVRI$, $UBVRIJHK$, and GALEX+$UBVRI$.

The structure of the paper is the following: in Section \ref{sec:grid_of_models} we describe our grid of star cluster models used to derive the physical parameters of unresolved star clusters. Section \ref{sec:method_cluster_parameters_determination} presents the method of deriving the cluster physical parameters developed in Paper I, but here with the addition of the metallicity dimension. Section \ref{sec:1Z_vs_1Z} discusses the ability of the method to derive the age, mass, and extinction, using different photometric systems. The input metallicity of the artificial cluster sample is fixed to one value and the parameters are derived using a model grid of one fixed metallicity, which can be different from the input metallicity. This allows inspecting the metallicity effect on the derivation of the age, mass, and extinction. Section \ref{sec:1Z_vs_3Z} explores the derivation of cluster parameters when the model grid contains models of different metallicities, so that we evaluate how accurately we can constrain the metallicity in addition to the age, mass, and extinction for various photometric systems. Conclusions are presented in Section \ref{sec:conclusions}.

\section{Model grid for age, mass, extinction, and metallicity}
\label{sec:grid_of_models}
To derive physical parameters of unresolved star clusters, we computed a large grid of discrete cluster models for age, mass, extinction, and metallicity, randomly sampling the stellar mass according to the initial mass function \citep[IMF,][]{Kroupa2001} following the method described in \citet{Deveikis2008} \citep[see also][]{Santos1997}. The star luminosities are derived from stellar isochrones of the selected age and metallicity of the cluster models. We used the PADOVA isochrones\footnote{PADOVA isochrones from ``CMD 2.5'': http://stev.oapd.inaf.it/cmd} from \citet{Marigo2008} with the corrections of \citet{Girardi2010} for the TP-AGB phases. The grid was built according to the following nodes: from $\log_{10}(t/\mathrm{yr})=6.6$ to 10.1 in steps of 0.05, from $\log_{10}(M/M_{\sun})=2$ to 5 in steps of 0.05, and for 13 metallicities: from $Z=0.03$ to $Z=0.00013$. This gives a grid of 71 values of age, 61 values of mass, with $1\,000$ models per node, hence $\sim$$4\times10^{6}$ models for each metallicity. To limit the number of models that need to be stored in computer memory, the extinction was computed when the observed cluster was compared with the grid of models. It ranges from $E(B-V)=0$ to 1 in steps of 0.01, therefore 101 values for the extinction. We used the Milky Way standard extinction law from \citet{Cardelli1989}.

\section{Deriving the cluster parameters}
\label{sec:method_cluster_parameters_determination}
The main idea of the method for determining the physical parameters (age, mass, extinction\footnote{We refer to extinction as $E(B-V)$ hereafter.}, and metallicity) of a given observed star cluster with available broad-band photometry is to compare its magnitudes with those of a 4D grid of models for every value of the four physical parameters. Each node of the model grid described in the previous section contains a large series of models of the same age, mass, extinction, and metallicity to represent stochastic variations in photometry.

In a similar way as \citet{Fouesneau2010} and \citet{Fouesneau2014}, we evaluated the likelihood of each node of the grid to represent the observed magnitudes of the cluster. For each node, we first computed the likelihood of each model from the node by
\begin{equation}
L_{\mathrm{model}} = \prod_{f=1}^{F} \frac{1}{ \sqrt{2 \pi}\, \sigma_{f}} \exp \left[ - \frac{\left(\mathrm{mag}_{f,\mathrm{obs}}-\mathrm{mag}_{f,\mathrm{model}}\right)^{2}}{2\,\sigma_{f}^{2}} \right]\,,
\end{equation}
where $f$ stands for one particular filter, $\mathrm{mag}_{f}$ for the observed and model magnitudes in that filter, and $F$ for the total number of filters, for example 5 for the $UBVRI$ photometric system. Then the likelihood of the node of age $t$, mass $M$, extinction $E(B-V)$, and metallicity $Z$ is the sum of the likelihoods of its models,
\begin{equation}
L_{\mathrm{node}}\left(t,M,E(B-V),Z\right) = \sum_{n=1}^{N} L_{\mathrm{model},\,n}\,,
\end{equation}
where $N$ is the total number of models contained in the node. The observed star cluster is then classified with the parameters of the node, which maximizes $L_{\mathrm{node}}$ among all nodes of the grid.

\section{Tests of the method on artificial cluster samples: one--$Z$ vs one--$Z$}
\label{sec:1Z_vs_1Z}
To characterize the accuracy of the method presented in Section \ref{sec:method_cluster_parameters_determination}, we built samples of $10\,000$ stochastic artificial clusters with a flat random distribution in the age range $\log_{10}(t/\mathrm{yr})=[6.6,10.1]$ and a flat random distribution in the extinction range $E(B-V)=[0,1]$, with the mass fixed to $\log_{10}(M/M_{\sun})=4$ and for 3 metallicities, $Z=0.019$, 0.005, and 0.00019. These artificial clusters were built with photometry available in optical ($UBVRI$), near-infrared ($JHK$), and ultraviolet (GALEX) filters.

\subsection{Parameter derivation for artificial clusters in $UBVRI$}
In this first test, we show the results of deriving age, mass, and extinction using $UBVRI$ filters alone. Photometric errors were included in the $UBVRI$ photometry of artificial clusters by adding Gaussian errors of $\sigma=0.05$ mag to each magnitude.

\begin{figure*}
\centering
\includegraphics[scale=0.55]{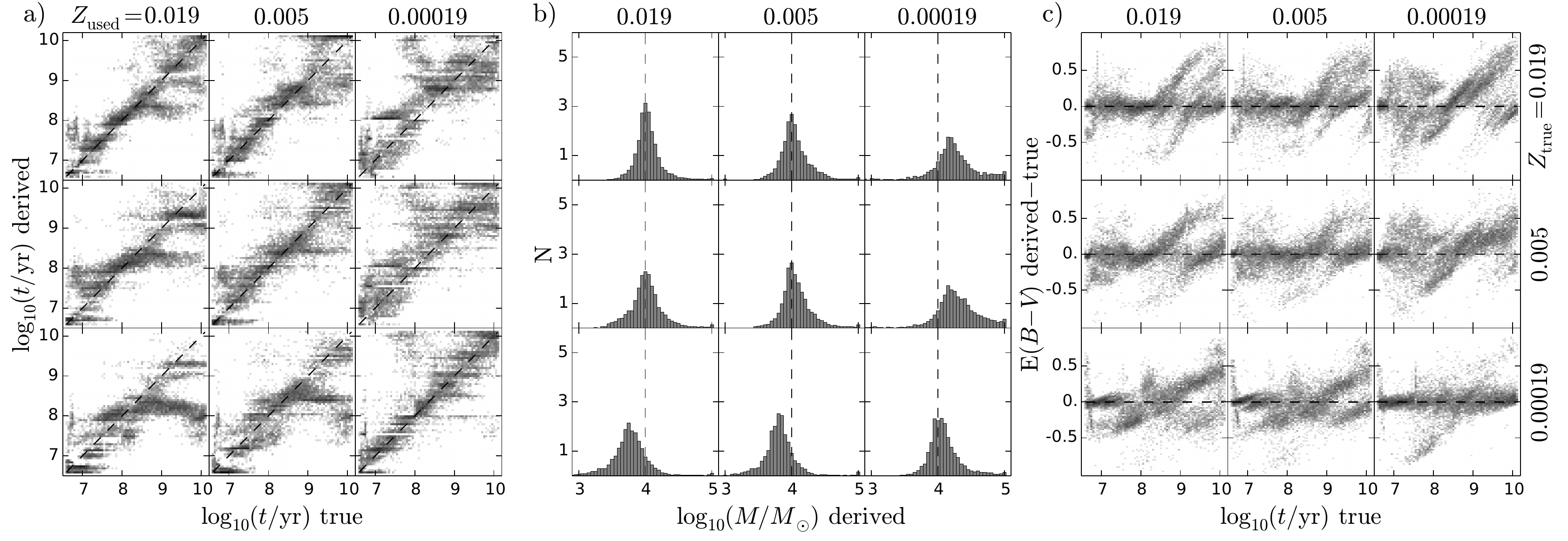}
\caption{Results of the test with nine samples of $10\,000$ artificial clusters using $UBVRI$ photometry with $\sigma=0.05$ mag for the photometric Gaussian errors for each filter. The true metallicity of the models is specified at the extreme right of the figure and the metallicity of the grid used to classify the clusters is specified at the top of panels. Blocks of panels show the derived \textbf{a)} age, \textbf{b)} mass (the total area of each histogram is normalized to one), and \textbf{c)} extinction. In each block, the main diagonal panels are those where the true metallicity and the metallicity of the grid are the same, $Z_{\mathrm{true}}=Z_{\mathrm{used}}$.}
\label{fig:1Z_vs_1Z_results_UBVRI}
\end{figure*}

Fig. \ref{fig:1Z_vs_1Z_results_UBVRI} shows the results, which show that the parameters are derived best when the metallicity used to classify clusters is the same as the true metallicity; see results in the main diagonal panels of the three panel blocks of Fig. \ref{fig:1Z_vs_1Z_results_UBVRI}. In the main diagonal panels of the age parameter (Fig. \ref{fig:1Z_vs_1Z_results_UBVRI} a), the streaks that develop perpendicularly up and down on the one-to-one black dashed line are caused by the age-extinction degeneracy. Old clusters with low intrinsic extinction can be mistakenly classified as younger clusters with higher extinction, and vice versa \citep[see ][as well as Paper I for more details on the age-extinction degeneracy]{Fouesneau2010}.

\begin{figure}
\centering
\includegraphics[scale=0.7]{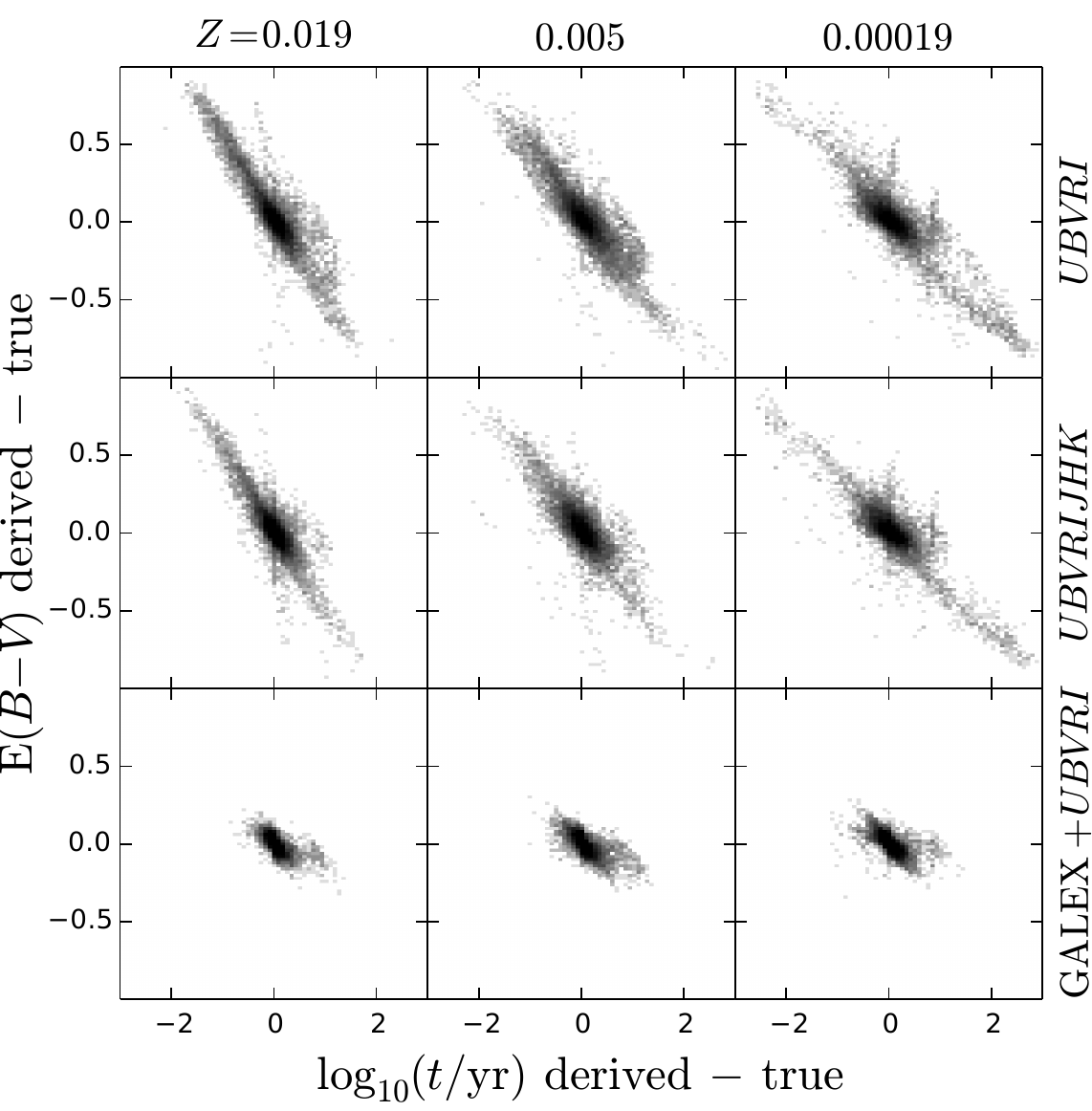}
\caption{Difference of the derived and the true extinction versus the difference of the derived and the true age for the sample of $10\,000$ artificial clusters. Clusters have been classified according to their true metallicity $(Z_{\mathrm{true}}=Z_{\mathrm{used}})$, which is indicated at the top of panels, and with different photometric systems, indicated at the right.}
\label{fig:1Z_vs_1Z_results_diffEbv_vs_diffAge}
\end{figure}

Fig. \ref{fig:1Z_vs_1Z_results_diffEbv_vs_diffAge} shows the difference of extinction (derived $-$ true) vs the difference of age (derived $-$ true) for clusters with true metallicity that were analyzed with model grids of the same metallicity, like in the main diagonal panels of Fig. \ref{fig:1Z_vs_1Z_results_UBVRI}. The first row of panels in Fig. \ref{fig:1Z_vs_1Z_results_diffEbv_vs_diffAge} shows results obtained with the $UBVRI$ photometric system. The extinction and age derivations are tightly related: an underestimation of the age correlates with an overestimation of the extinction, and an overestimation of the age correlates with an underestimation of the extinction. With decreasing metallicity, we note a slight change of the degeneracy correlation and concentration.

The panels above and below the main diagonal panels of Fig. \ref{fig:1Z_vs_1Z_results_UBVRI} show results where the metallicity of the grid is different from the true one, therefore additional biases appear. For the age parameter we observe a reinforcement of the left part of the age-extinction degeneracy streak when the grid metallicity is lower than the true metallicity (panels above the main diagonal panels) and a reinforcement of the right part of the streak when grid metallicity is higher than true metallicity.

Essentially when the grid metallicity is higher than the true metallicity, the underestimation of the ages of old clusters can be understood because, for the same age, mass, and extinction, low-metallicity clusters are generally brighter and bluer than high-metallicity clusters. This is why classifying low-metallicity clusters with a higher-metallicity grid will interpret this brightness as that of younger clusters. Moreover, the degeneracy exists not only between the age and metallicity: the clusters are perceived to be much younger than they are, but also less massive (Fig. \ref{fig:1Z_vs_1Z_results_UBVRI} b). The extinction (Fig. \ref{fig:1Z_vs_1Z_results_UBVRI} c) is also affected, so that the degeneracies are strongly multidimensional. The reverse effects are also true for the upper main diagonal panels, when the grid metallicity is lower than the true one. Out of main diagonal results show the biases introduced when cluster parameters are derived using a model grid with an incorrect metallicity.

\subsection{Addition of near-infrared passbands: $UBVRI$+$JHK$}

\begin{figure*}
\centering
\includegraphics[scale=0.55]{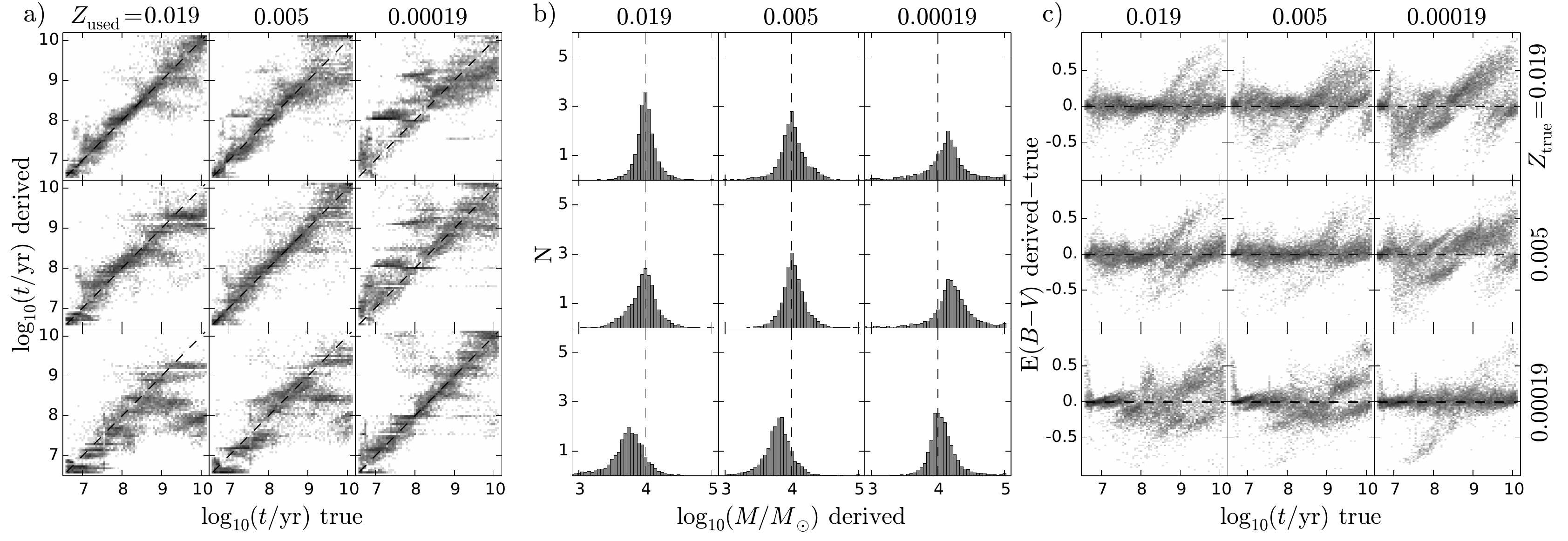}
\caption{Same as Fig. \ref{fig:1Z_vs_1Z_results_UBVRI}, but for $UBVRIJHK$ filters with $\sigma=0.05$ mag for the photometric Gaussian errors for each $UBVRI$ filter, and $\sigma=0.1$ mag for each $JHK$ filter.}
\label{fig:1Z_vs_1Z_results_UBVRIJHK}
\end{figure*}

By using simple stellar population models (SSP), \citet{Bridzius2008} \citep[see also][]{Anders2004} showed that adding near-infrared information ($JHK$ photometric bands) to optical $UBVRI$ filters could help to significantly decrease the age-extinction degeneracy. We performed the same tests with stochastic clusters with additional $JHK$ filters. In each $JHK$ filter, we added Gaussian photometric errors with $\sigma=0.1$ mag to mimic the larger photometric uncertainties present in these passbands. We kept $\sigma=0.05$ mag for the Gaussian photometric errors added to the $UBVRI$ filters.

The results are displayed in Fig. \ref{fig:1Z_vs_1Z_results_UBVRIJHK}. The main diagonal panels of the age and extinction parameter blocks show that the age-extinction degeneracy is reduced significantly. The streaks perpendicular to the one-to-one red line in the age parameter block (Fig. \ref{fig:1Z_vs_1Z_results_UBVRIJHK} a) are much less populated than when only $UBVRI$ filters were used (Fig. \ref{fig:1Z_vs_1Z_results_UBVRI} a). The decrease of intensity of the age-extinction degeneracy is also visible in Fig. \ref{fig:1Z_vs_1Z_results_diffEbv_vs_diffAge} (second row), compared with the case without $JHK$ filters (first row).

However, out of the main diagonal panels of Fig. \ref{fig:1Z_vs_1Z_results_UBVRIJHK}, when the grid metallicity is different from the true metallicity of clusters, the strong additional biases due to metallicity observed in $UBVRI$-only case are still present, despite some minor differences.

\subsection{Addition of ultraviolet passbands: GALEX+$UBVRI$}

\begin{figure*}
\centering
\includegraphics[scale=0.55]{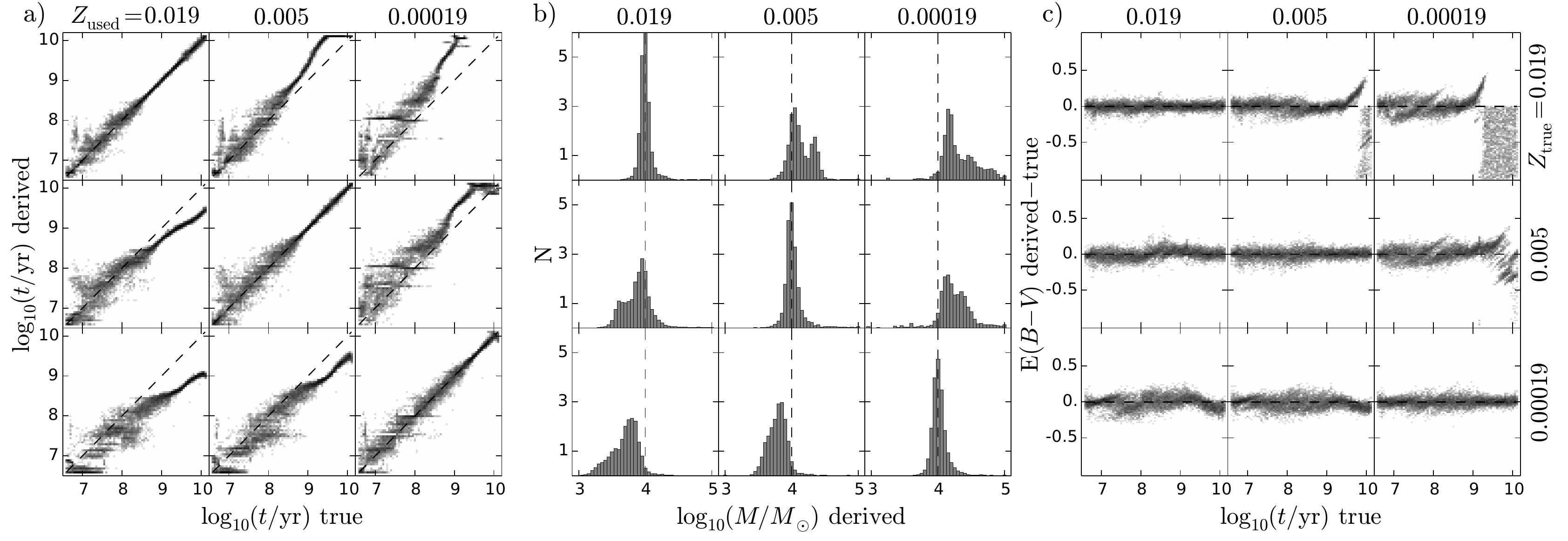}
\caption{Same as Fig. \ref{fig:1Z_vs_1Z_results_UBVRI}, but with GALEX (FUV and NUV) and $UBVRI$ filters with $\sigma=0.05$ mag for the photometric Gaussian errors for each $UBVRI$ filter, and $\sigma=0.15$ mag for each GALEX filter. The shaded areas in the panels of the extinction parameter in block \textbf{c)}, which range from 0 to $-1$ indicate that there are no solutions in these cases.}
\label{fig:1Z_vs_1Z_results_GALEX_UBVRI}
\end{figure*}

It is well-known \citep[e.g., ][]{Kaviraj2007} that the association of far-ultraviolet (FUV, $\lambda_{\mathrm{eff}}=1539\AA$) and near-ultraviolet (NUV, $\lambda_{\mathrm{eff}}=2316\AA$) passbands with optical $UBVRI$ passbands helps in reducing degeneracies in the derivation of parameters. Here we study the $10\,000$ stochastic artificial star clusters with $UBVRI$ photometry, with $\sigma=0.05$ mag for the photometric Gaussian errors, and GALEX photometry with $\sigma=0.15$ mag for each of the FUV and NUV filters.

Fig. \ref{fig:1Z_vs_1Z_results_GALEX_UBVRI} shows the results for this filter combination, sharply constraining the age, mass, and extinction. The age-extinction streaks present in main diagonal panels in previous tests (Fig. \ref{fig:1Z_vs_1Z_results_UBVRI} and \ref{fig:1Z_vs_1Z_results_UBVRIJHK}) here vanish entirely. This is also clear in the last row of Fig. \ref{fig:1Z_vs_1Z_results_diffEbv_vs_diffAge}, where the correlation present in $UBVRI$ and $UBVRIJHK$ fades.

The break of the age-extinction degeneracy brought by ultraviolet data can be understood in Fig. \ref{fig:GALEX_SSP}. Panel (a) shows the SSPs of the 13 metallicities in optical passbands alone as well as the distributions of $1\,000$ discrete cluster models of mass $\log(M/M_{\sun})=4$, of age $\log_{10}(t/\mathrm{yr})=7$, 7.5, 8, 8.5, 9, 9.5, and 10, for three metallicities: $Z=0.019$, 0.005, and 0.00019. In that case, for a given SSP of fixed metallicity, the bending of the SSP curve at intermediary ages and the direction of the reddening vector indicate an age-extinction degeneracy between young and old clusters. This is also true in terms of discrete model distributions, because the distributions of a given metallicity follow the SSP line of the corresponding metallicity. Fig. \ref{fig:GALEX_SSP} (b) shows that this is not anymore the case when ultraviolet data are used, as the reddening vector now strongly deviates from the SSP direction.

However, as seen in previous tests, offsets are also observed in Fig. \ref{fig:1Z_vs_1Z_results_GALEX_UBVRI} (a) when the grid metallicity is different from the true one (out of main diagonal panels), but the scatter is much smaller here. As before, when the metallicity of the grid is higher than the true one (panels below the main diagonal panels), the age is classified as younger and the mass as lower, and the reverse is true when the grid metallicity is lower than the true one (panels above main diagonal panels). 

Note, however, that these results are in practice not useful for clusters older than a few hundred million years, as the FUV and NUV magnitudes strongly fade later on \citep[see, e.g.,][]{Bianchi2011}. This means that to estimate the parameters of star clusters of close galaxies such as Andromeda or Triangulum, GALEX data can only be used for young clusters because for older clusters ultraviolet photometry falls below detection limit.

\section{Tests of the method on artificial cluster samples: one--$Z$ vs whole metallicity range}
\label{sec:1Z_vs_3Z}

\begin{figure*}
\centering
\includegraphics[scale=0.489]{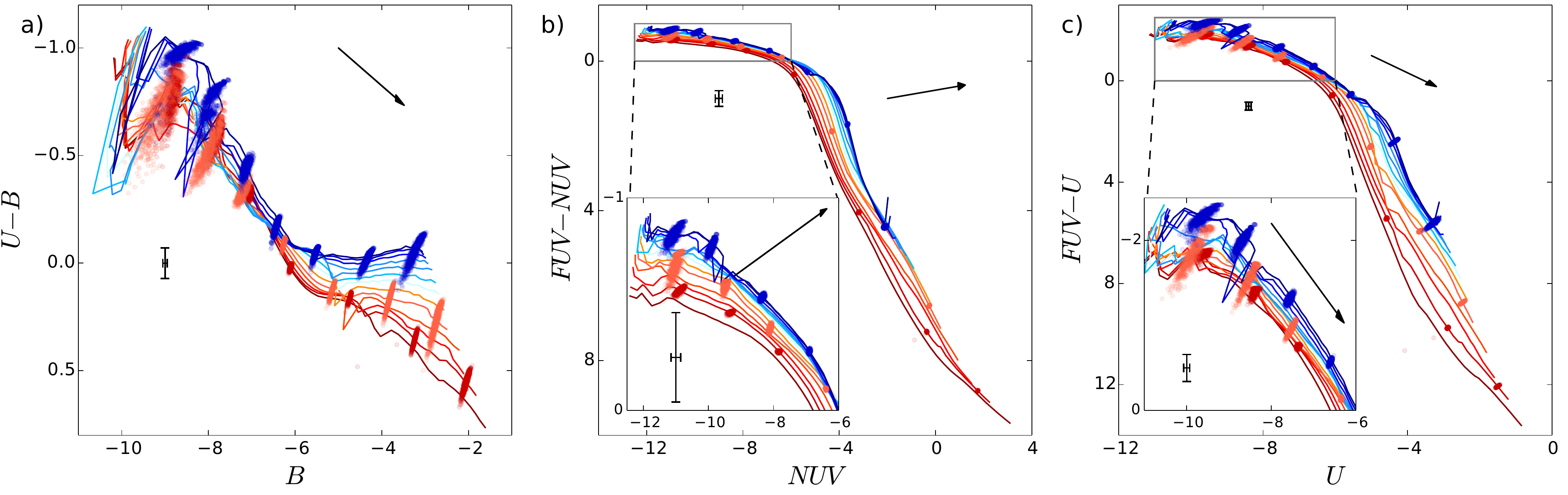}
\caption{Color vs magnitude diagrams showing the SSP curves of the 13 metallicities used in the model grid, with discrete model distributions of age $\log_{10}(t/\mathrm{yr})=7$, 7.5, 8, 8.5, 9, 9.5, and 10 (young models are top left and old models are bottom right in each panel), for three metallicities: $Z=0.019$ (red), 0.005 (orange) and 0.00019 (blue). Each distribution contains $1\,000$ discrete cluster models with mass fixed to $\log(M/M_{\sun})=4$, and the SSPs are scaled to the mass of the clusters. The reddest SSP line indicates highest metallicity $Z=0.03$ and the bluest SSP line indicates lowest metallicity $Z=0.00013$. The 1--$\sigma$ ($\sigma=0.15$ mag for FUV and NUV, and $0.05$ mag for $U$ and $B$ passbands) error bars show the standard deviations of the photometric accuracy. The black arrows indicate the extinction vector, computed for $A_{V}=1$ mag. Panel \textbf{a)} shows the situation in optical bands alone, while in panels \textbf{b)} and \textbf{c)}, ultraviolet information is present, and the zoom plots show details at young age.}
\label{fig:GALEX_SSP}
\end{figure*}

\begin{figure*}
\centering
\includegraphics[scale=0.62]{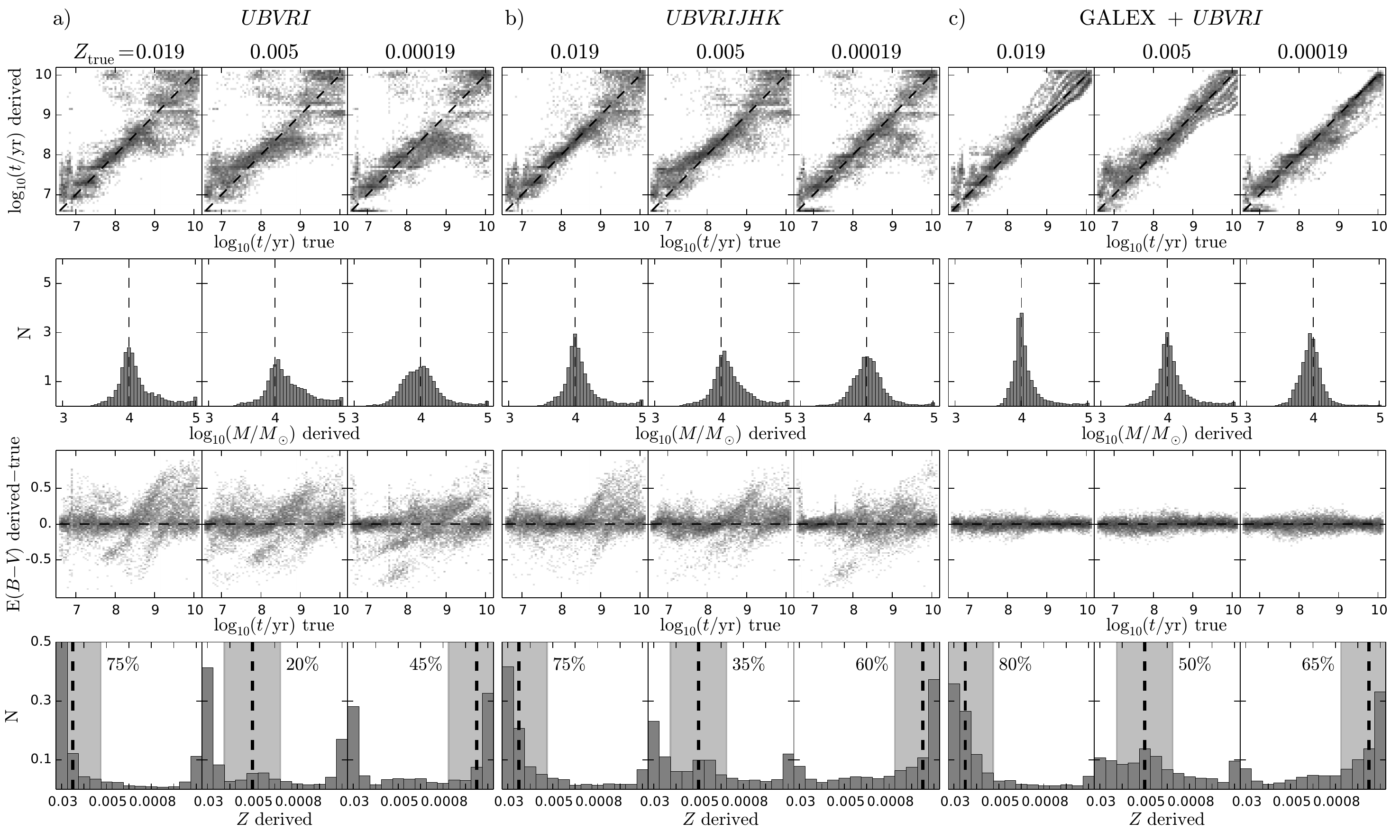}
\caption{Age, mass, extinction, and metallicity derived for samples of $10\,000$ artificial clusters and with true mass $\log_{10}(M/M_{\sun})=4$ when the metallicity is unknown. In each case, the true metallicity is indicated at the top of each column, and also by the vertical dashed line in metallicity panels (bottom row of panels). In the block of panels \textbf{a)}, only $UBVRI$ photometry has been used with $\sigma=0.05$ mag of photometric errors. In block \textbf{b)} $JHK$ filters have been added to $UBVRI$, with $\sigma=0.1$ mag of photometric errors. In block \textbf{c)} GALEX filters have been added to $UBVRI$, with 0.15 mag of photometric errors. In metallicity panels, regions are defined around the true metallicities (shaded area $\pm$2 bins around true metallicities). The percentage of clusters classified according to metallicity in this region is indicated. Mass and metallicity histograms are normalized so that the area below the histogram is 1.}
\label{fig:1Z_vs_13Z_results_Ebv}
\end{figure*}

\begin{figure*}
\centering
\includegraphics[scale=0.62]{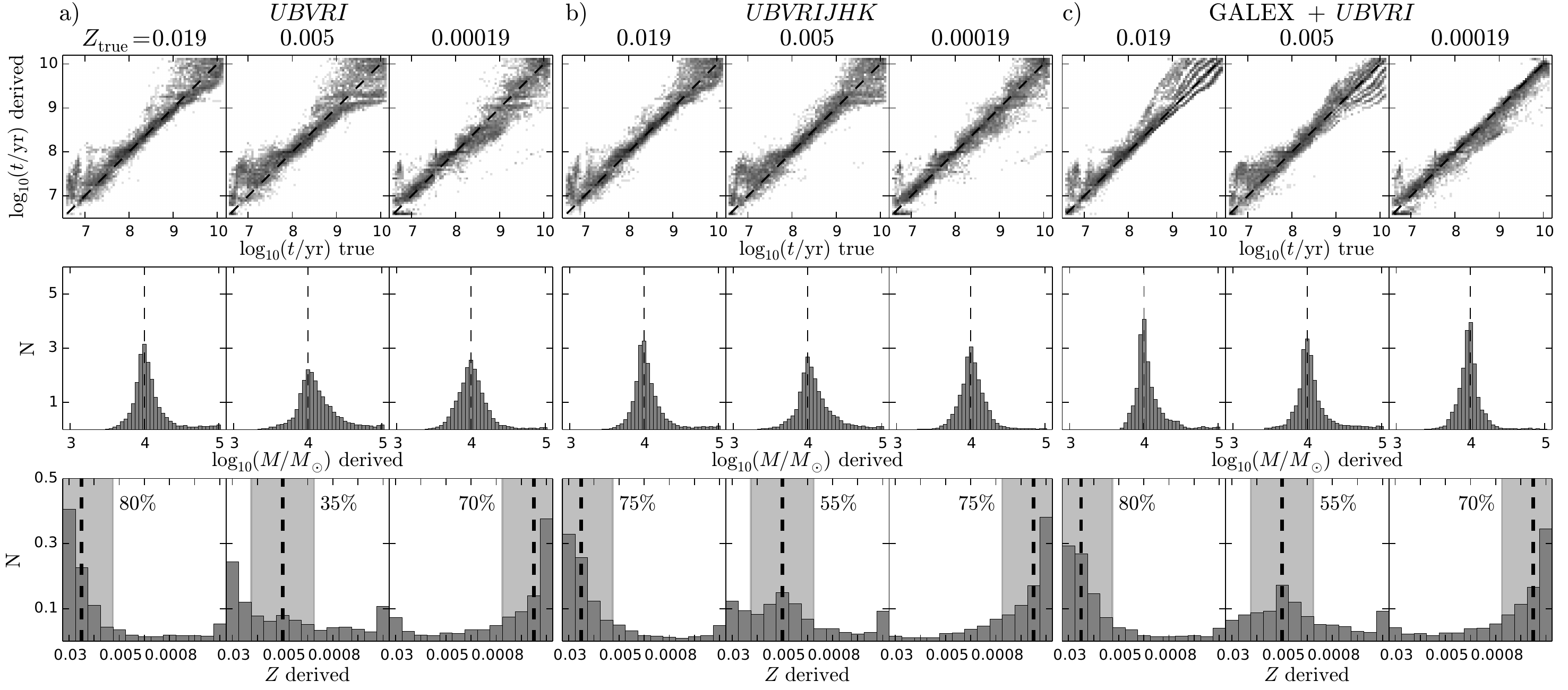}
\caption{Same as Fig. \ref{fig:1Z_vs_13Z_results_Ebv}, but for star cluster models without extinction added to their photometry, $E(B-V)=0$.}
\label{fig:1Z_vs_13Z_results_noEbv}
\end{figure*}

In this section, we study the parameter derivation of three artificial clusters samples, each of them with one fixed metallicity ($Z=0.019$, 0.005, and 0.00019) using a large grid containing cluster models of 13 different metallicities (from $Z=0.03$ to $Z=0.00013$), to determine the ability of the method to derive the cluster age, mass, and extinction when the metallicity is unknown, with an attempt to also constrain the metallicity itself. First, we apply the method using only $UBVRI$ photometry, then using $UBVRIJHK$ photometry, and finally using GALEX+$UBVRI$ photometry. As previously, the mass of clusters was fixed to $\log(M/M_{\sun})=4$, and the models were randomly reddened in the range $E(B-V)=[0,1]$. As before, the photometric errors for $UBVRI$, $JHK$ and GALEX filters were $\sigma=0.05$, 0.1, and 0.15 mag, respectively.

Fig. \ref{fig:1Z_vs_13Z_results_Ebv}(a) shows the results obtained for clusters studied using $UBVRI$ photometry. Concerning the age, we still see the age-extinction degeneracy streaks perpendicular to the one-to-one black dashed line, which extends differently depending on the true metallicity of clusters.

In Fig. \ref{fig:GALEX_SSP}(a), the age-extinction degeneracy exists for each given SSP with fixed metallicity. When we derive the cluster parameters allowing all metallicities, we see that the age-extinction degeneracy now extends between SSPs (and also between the discrete model distributions) of different metallicities, enlarging the uncertainty in deriving these parameters.

In addition to the reinforced age-extinction degeneracy, deriving the metallicity is complicated by the fact that in optical filters the SSPs (or the discrete model distributions of corresponding age) are naturally close and overlapping for young and intermediate-age clusters, so that the distance between them in photometric space is small compared with typical photometric uncertainties. These two problems make the $UBVRI$ photometric system unsuitable in deriving the metallicity.

In Fig. \ref{fig:1Z_vs_13Z_results_Ebv}, we define regions around the true metallicities (shaded area $\pm$2 bins around the vertical line in bottom panels) in which we quantify how many solutions are found in these regions. For example, for clusters with true metallicity $Z=0.005$ (central case of Fig. \ref{fig:1Z_vs_13Z_results_Ebv} a), only $\sim$20\% of clusters are classified with a metallicity in the region around the true metallicity when only $UBVRI$ is used. The accumulation of solutions at the metallicity range boundaries can be interpreted as that the derived metallicity of these clusters could be located even farther away from the true values, if the range of metallicities were broader.

Fig. \ref{fig:1Z_vs_13Z_results_Ebv} (b) shows the results of the same clusters, now studied with $UBVRIJHK$ photometry. The change compared with the $UBVRI$ case is the reduction of the dispersion in age and extinction panels. Concerning the metallicity derivation, the clusters are still suffering from metallicity boundary effects, but the effect is reduced, the metallicity is derived slightly better. For example, $\sim$35\% of clusters with true metallicity $Z=0.005$ are classified with a metallicity within the good region. And $\sim$60\% of clusters with true metallicity $Z=0.00019$ are classified correctly, much better than in the case with $UBVRI$ alone. In that case, the strong degeneracy with boundary metallicity $Z=0.03$ almost vanishes.

Fig. \ref{fig:1Z_vs_13Z_results_Ebv} (c) shows the results of the same clusters, now studied with GALEX+$UBVRI$ photometry. Now derived and true parameters agree much better. As in the one--$Z$ vs one--$Z$ tests of the previous section, the age panels show that adding ultraviolet information completely discards the streaks perpendicular to the one-to-one black dashed line. Fig. \ref{fig:GALEX_SSP}(b) shows that young clusters cannot be degenerated with old clusters through age-extinction degeneracy, which facilitates deriving the age and mass. The scatter around the true mass is also more strongly reduced than for $UBVRI$ and $UBVRIJHK$. The metallicities are now better derived as well, even when the true metallicity is $Z=0.005$, the boundary effects are significantly reduced. Moreover, at least half of the clusters are classified with the correct metallicity when ultraviolet data are added to the optical. In Fig. \ref{fig:GALEX_SSP}(b) and (c), the SSPs (and also the discrete model distributions) overlap much less than in optical passbands. In Fig. \ref{fig:GALEX_SSP}(c), the combination of far-ultraviolet with optical $U$ passband allows us to distinguish between the different metallicities, as the reddening vector is now nearly parallel to the SSPs (especially at young ages) and that the SSPs are more spaced, compared with the error bars (see the zoom plot). It is also interesting to note that because the reddening vector points in different directions in Figs. \ref{fig:GALEX_SSP}(b) and (c), it is not possible to degenerate two discrete model distributions of different metallicities by means of extinction.

Figures \ref{fig:GALEX_SSP} and \ref{fig:1Z_vs_13Z_results_Ebv} (c) are reproduced in the appendix (Figs. \ref{fig:GALEX_SSP_LOGM3} and \ref{fig:1Z_vs_13Z_results_noEbv_LOGM3}) for lower-mass star clusters ($\log_{10}(M/M_{\sun})=3$) to show that the method is also applicable for lower-mass clusters.

In Fig. \ref{fig:1Z_vs_13Z_results_noEbv} the same tests were performed with artificial clusters \emph{without} any extinction added to their photometry ($E(B-V)=0$), and classified considering the extinction as a known parameter, leaving only the age, mass, and metallicity as free parameters. Now, without degeneracies introduced by extinction, the derivation of the other parameters is much more accurate. For all photometric systems, the effect of the metallicity on the age derivation manifests itself through a slight shift of the ages above or below the one-to-one black dashed line. Concerning the metallicity derivation, one can see that even in this case when there is no extinction, a consistent derivation of the metallicity with only $UBVRI$ filters is not possible, despite a strong decrease of boundary effects on the metallicity, compared with the case where extinction was included (compare Figs. \ref{fig:1Z_vs_13Z_results_Ebv} and \ref{fig:1Z_vs_13Z_results_noEbv}). For example, only $\sim$35\% of clusters are classified in the region of the correct metallicity when the true metallicity is $Z=0.005$. Using $UBVRIJHK$, the metallicity of clusters is better derived than in the case with unknown extinction, and the metallicity boundary effects are decreased. In the worst case, at least $\sim$55\% of clusters are classified with a correct metallicity. Hence, the break of degeneracies with metallicity can be efficiently achieved with $UBVRIJHK$ photometric system only when the extinction is well constrained beforehand. For GALEX+$UBVRI$, the metallicity prediction only slightly improves when there is no extinction, compared with the case when there is extinction.

\section{Conclusions}
\label{sec:conclusions}
We have completed the task started in Paper I, which aims to derive the main physical parameters of unresolved star clusters using broad-band photometry when the stochastic sampling of stellar mass in clusters is taken into account. The influence of the metallicity was studied in this paper.

We showed the biases induced by metallicity when artificial clusters are classified with a grid of models of a different metallicity. We found that optical ($UBVRI$) photometric data provide strongly biased results, which can be reduced when near-infrared ($JHK$) data are added. Adding ultraviolet (GALEX) data allows strongly constraining the age, mass, and extinction parameters. Deviations appear for old clusters, but the fading of ultraviolet photometry below the detection limit after a few hundred million years makes GALEX data unsuitable for old clusters in close galaxies such as Andromeda and Triangulum.

We tested the ability of optical, optical+near-infrared, and optical+ultraviolet photometric systems of constraining the metallicity using samples of clusters of fixed true metallicity, and classified with the grids of models with 13 metallicities ranging from $Z=0.03$ to $Z=0.00013$. We conclude that deriving the metallicity using broad-band photometry for unresolved star clusters requires a wide spectral information including ultraviolet passbands with errors as small as 0.05 mag in the optical and 0.15 mag in the ultraviolet. Finally, we also demonstrated that knowing the extinction helps constraining the age of clusters and slightly improves the metallicity prediction. In that case, optical associated with near-infrared data can significantly reduce the biases that affect the metallicity prediction.

\begin{acknowledgements}
This research was funded by a grant (No. MIP-074/2013) from the Research Council of Lithuania. We thank the anonymous referee for comments and suggestions that significantly improved the paper.
\end{acknowledgements}

\bibliographystyle{aa}
\bibliography{Article_II_Metallicity}

\section{Appendix}
\label{sec:appendix}

\begin{figure*}
\centering
\includegraphics[scale=0.489]{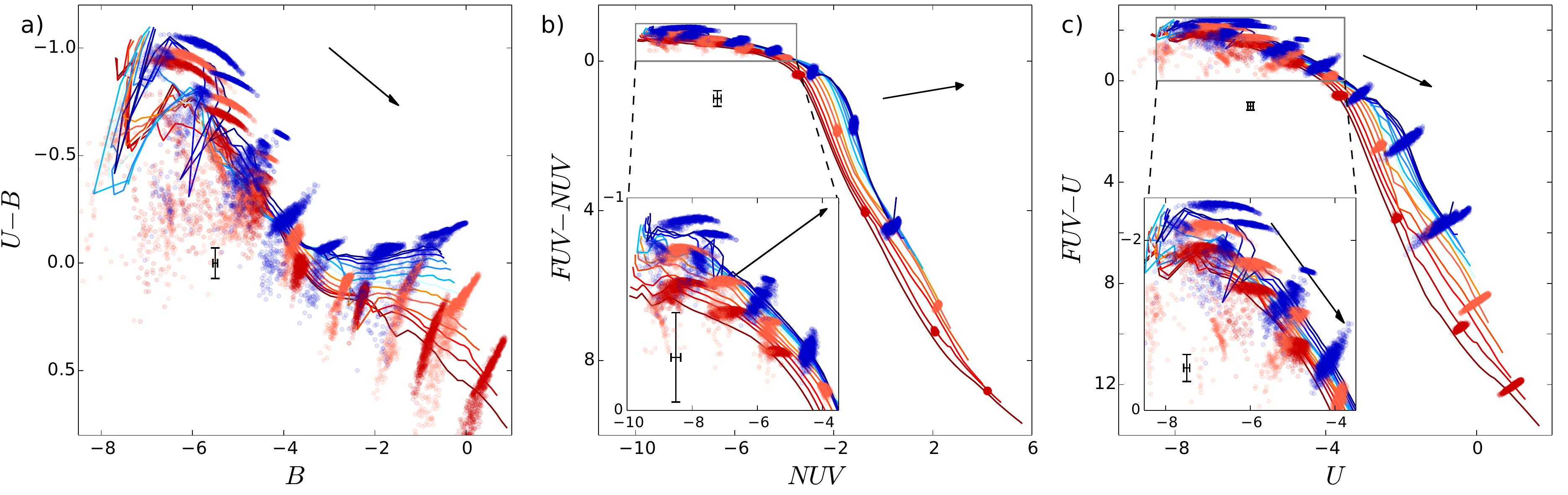}
\caption{Same as Fig. \ref{fig:GALEX_SSP}, but for star cluster discrete models with mass $\log_{10}(M/M_{\odot})=3$. The SSPs are scaled to the mass of the clusters.}
\label{fig:GALEX_SSP_LOGM3}
\end{figure*}

\begin{figure}
\centering
\includegraphics[scale=0.7]{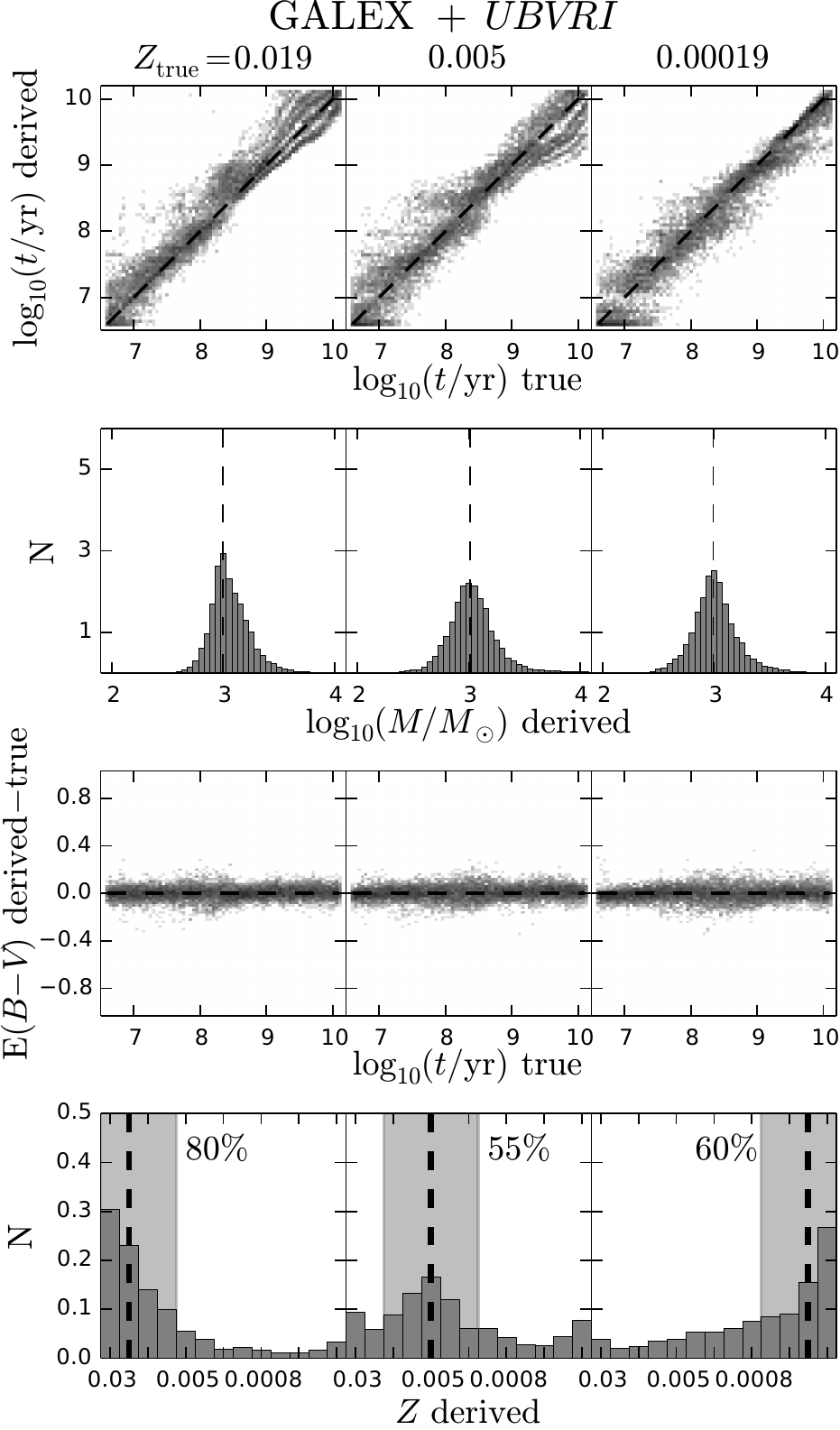}
\caption{Same as Fig. \ref{fig:1Z_vs_13Z_results_Ebv}, but for star cluster models with mass $\log_{10}(M/M_{\odot})=3$ for the GALEX+$UBVRI$ photometric system.}
\label{fig:1Z_vs_13Z_results_noEbv_LOGM3}
\end{figure}

Here we derive the age, mass, extinction, and metallicity of an artificial cluster sample in which the mass of clusters is fixed to $\log_{10}(M/M_{\sun})=3$, to show that the method of parameter derivation can be applied for low-mass clusters.

For this purpose, Fig. \ref{fig:GALEX_SSP_LOGM3} displays discrete models of age $\log_{10}(t/\mathrm{yr})=7$, 7.5, 8, 8.5, 9, 9.5, and 10 for three metallicities: $Z=0.019$, 0.005 and 0.00019. The SSPs are also shown for the 13 metallicities used in the paper. While the dispersion for these discrete models is wider than for the case of models $\log_{10}(M/M_{\sun})=4$ (see Fig. \ref{fig:GALEX_SSP}), the general configuration of the models remains the same. Hence, while the optical-only color-magnitude diagram (Fig. \ref{fig:GALEX_SSP_LOGM3} a) still shows strong age-extinction degeneracy, we see that, as in the case of more massive models, including ultraviolet information allows breaking the age-extinction degeneracy (Fig. \ref{fig:GALEX_SSP_LOGM3} b). Although the wings of discrete model distributions are more elongated than in the $\log_{10}(M/M_{\sun})=4$ case, the cores of each distribution are still rather well separated from distributions of different metallicities (Figs. \ref{fig:GALEX_SSP_LOGM3} b and c). In addition, as the reddening vector points in different directions in Figs. \ref{fig:GALEX_SSP_LOGM3} (b) and (c), it is not possible to degenerate these distributions by means of extinction.

The results of deriving the parameters of the artificial cluster sample are displayed in Fig. \ref{fig:1Z_vs_13Z_results_noEbv_LOGM3}, for GALEX+$UBVRI$. The break of the age-extinction degeneracy translates into a narrow scatter around the true values of these parameters (Fig. \ref{fig:1Z_vs_13Z_results_noEbv_LOGM3}, first and third row for the age and extinction, respectively). The reduced degree of freedom of the extinction also allows a satisfactory derivation of the mass and metallicity (Fig. \ref{fig:1Z_vs_13Z_results_noEbv_LOGM3}, second and last rows), although a larger scatter around the true mass is observed in the $\log_{10}(M/M_{\sun})=3$ case than for $\log_{10}(M/M_{\sun})=4$.

\end{document}